\documentclass[11pt,twoside]{book}
\usepackage{konkolyproc2}
\usepackage{longtable}
\usepackage{amsmath,amssymb}
\usepackage{graphicx}
\usepackage{lscape}
\usepackage{index}
\usepackage{natbib}
\usepackage{bigdelim}
\usepackage{multirow}
\makeindex

\begin{document}

\pagestyle{myheadings}
\setcounter{equation}{0}\setcounter{figure}{0}\setcounter{footnote}{0}
\setcounter{section}{0}\setcounter{table}{0}\setcounter{page}{1}
\markboth{B\'anyai et al.}{Constraining RRc candidates using SDSS colours}
\title{Constraining RRc candidates using SDSS colours}
\author{E. B\'anyai$^1$, E. Plachy$^2$, L. Moln\'ar$^2$, L. Dobos$^1$ \&
R.~Szab\'o$^2$}
\affil{$^1$Dept. of Physics of Complex Systems, E\"otv\"os Lor\'and University, P\'azm\'any P. s\'et\'any 1/A, Budapest, 1117, Hungary\\
$^2$Konkoly Observatory, Research Centre for Astronomy and Earth Sciences, Hungarian Academy of Sciences, H-1121, Budapest, Konkoly Thege Mikl\'os \'ut 15-17, Hungary}

\begin{abstract}
The light variations of first-overtone RR Lyrae stars and contact eclipsing 
binaries can be difficult to distinguish. The Catalina Periodic Variable Star 
catalog contains several misclassified objects, despite the classification 
efforts by \citet{Drakeetal2014}. They used metallicity and surface gravity 
derived from spectroscopic data (from the SDSS database) to rule out binaries. 
Our aim is to further constrain the catalog using SDSS colours to estimate 
physical parameters for stars that did not have spectroscopic data.
\end{abstract}

\vspace*{-4mm}
\section{Method and data}

Briefly, \textsc{Photo-Met} estimates the unknown physical parameters of stars 
by interpolating the known parameters of other stars that have very similar 
broad-band colours. The method relies on two numerical algorithms: \\
\noindent - efficient $k$-nearest-neighbour finding in a four-dimensional metric 
colour-colour space and \\
\noindent - local linear regression.

As with any empirical parameter estimation algorithm, the reliability of the 
entire process depends much more on the training set than on the actual 
numerical method. For this study, we used an empirical training set based on 
SDSS PSF magnitudes of approximately 360\,000 stars. The stellar parameters [Fe/H], 
$T_{\rm eff}$ and $\log g$ are the adopted weighted averages from SSPP (SEGUE 
Stellar Parameter Pipeline). For further details on the method see 
\citet{Kerekesetal2013}.

The variable list of the Catalina Sky Survey consists of 5467 stars marked as 
RRc. The cross-match with the SDSS DR10 Cross-ID tool using an 1\farcs2 radius 
search resulted in 2762 stars. 
For delivering new results we took the stars with photometric measurements only 
(1732 objects) and applied the \textsc{Photo-Met} method to estimate the 
surface gravity, the effective temperature and metallicity.

\section{Constraining candidates}
To distinguish between RRc and contact binary stars we applied two criteria:\\
\noindent - RRc stars are halo giants so they are generally expected to have lower 
metallicity ([Fe/H] $< -1$) and surface gravity ($\log g < 3.6$) than disk
stars.\\
\noindent - Furthermore according to the work of \citet{Drakeetal2014} the RRc stars 
are concentrated between $ 0.24 < P < 0.42$ and $0.45 < M_{\rm test} < 0.55$ on 
the $M_{\rm test}-P$ plane, where $M_{\rm test}$ statistic values is a measure
of the fraction of time that an object spends below the mean magnitude.
We find 236 objects satisfied both, but 636 objects satisfied none of the criteria.

\section{Conclusion}
We could identify several contact binaries that were originally classified as 
RRc stars in the CSS Periodic Variable Star Catalog. The \textsc{Photo-Met} 
method, however, clearly placed them outside the RRc domain set by 
\citet{Drakeetal2014}. We plotted in Fig.~1 four light curves where the 
asymmetry of the minima can be clearly seen when folded with twice the assumed 
variation period. 
However, despite our success in finding new contaminating binary stars in the 
sample, the large errors in the $\log g$ and [Fe/H] determination (up to 
$\pm 1.0$ dex) and the low quality of the light curves of faint stars make 
this method uncertain, leaving a lot of ambiguous objects in the catalog. 

\begin{figure}[!ht]
\includegraphics[width=0.95\textwidth]{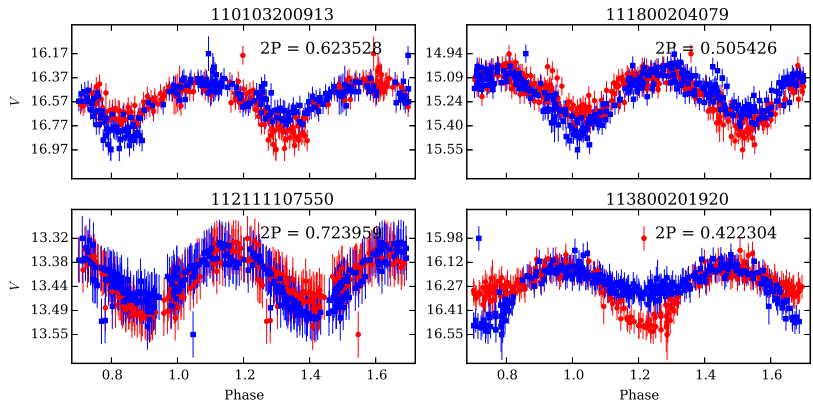}
\vspace*{-3mm}
\caption{Folded light curves of newly identified eclipsing binaries in the 
CSS RRc sample. Blue and red points are shifted in phase by 0.5 with respect 
to each other to highlight the differing minima.} 
\label{authorsurname-fig1} 
\end{figure}

\vspace*{-3mm}
\section*{Acknowledgements}
\vspace*{-1mm}This project has been supported 
by the LP2014-17 Program of the Hungarian Academy of Sciences, the NKFIH 
K-115709 and the OTKA NN-114560 grants of the Hungarian National Research, 
Development and Innovation Office. L.M. was supported by the J\'anos Bolyai 
Research Scholarship of the Hungarian Academy of Sciences. The CSS survey is 
funded by the National Aeronautics and Space Administration under Grant 
No. NNG05GF22G issued through the Science Mission Directorate Near-Earth 
Objects Observations Program.

\end{document}